\documentstyle[11pt,aaspp,psfig]{article}

\begin{document}
\title{Cosmological versus Intrinsic:
The Correlation between
Intensity and the Peak of the $\nu F_{\nu}$ Spectrum of
Gamma Ray Bursts.}
\author{Nicole M. Lloyd$^{1}$, Vah{\'e} Petrosian$^{1}$, 
Robert S. Mallozzi$^{2}$}
\affil{$^{1}$ Center for Space Sciences and Astrophysics, 
Stanford University, $^{2}$ Dept. of Physics, University of Alabama
in Huntsville.}

\begin{abstract}
We present results of correlation studies, examining
the association between the peak of the $\nu F_{\nu}$ spectrum
of gamma ray bursts, 
$E_{p}$,  with the burst's energy fluence and photon peak flux.
We discuss methods to account
for data truncation in $E_{p}$ and fluence or flux when performing
the correlation analyses.
 However, because bursts near the detector threshold are not
usually able to provide reliable spectral parameters, we focus on
 results
for the brightest bursts in which we can better understand
the selection effects relevant to
$E_{p}$ and burst strength. 
 We find that there is a strong
correlation between total fluence and $E_{p}$. 
We discuss these results
in terms of both cosmological and intrinsic effects.
  In particular, we show that for realistic distributions
of the burst parameters, cosmological expansion alone cannot account for the
correlation between $E_{p}$ and total fluence; the observed correlation
is likely a result of an intrinsic relation between the burst rest-frame
peak energy and the total radiated energy.  We investigate this latter
scenario in the context of synchrotron radiation from
external and internal shock models of GRBs. We find that the internal
shock model 
is consistent with our interpretation of the correlation, while the
external shock model 
cannot easily explain this intrinsic relation between peak energy and
burst radiated energy. 

{\it Subject headings}: gamma rays: bursts -- 
 cosmology: miscellaneous. 
\end{abstract}

\section{Introduction} 
 
   We now have confirmed redshift measurements to 
 eight gamma-ray bursts (GRBs): 970228
   (Djorgovski et al., 1999b), 970508 (Metzger et al., 1997, 
   Bloom et al., 1998), 
 971214 (Kulkarni et al., 1998), 
 980613 (Djorgovski et al., 1999a), 980703 (Djorgovski et al., 1998),  
 990123 (Kelson et al., 1999, Hjorth et al. 1999), 990510 (Vreeswijk
et al., 1999), 990712 (Galama et al., 1999).
  Although these redshifts have helped provide insight into
the energetics of GRBs, they have not helped us gain knowledge on
the cosmological distribution
of GRBs as a whole.  This is primarily because the ``luminosity
functions'' of these bursts are broad, and mask
any cosmological signature.  Figures 1a and 1b show the Hubble diagram for
GRBs and their afterglows with known redshifts  (and for which the
fluence and flux data is publicly available).
 There is no
obvious Hubble relation in any wavelength band, which suggests that
 we must  
continue to rely on statistical
studies of GRBs to gain insight on their global
distribution as well as the physical processes at the
burst.
 Log$N$-Log$S$ studies have been extensively used to
investigate the spatial distribution of GRBs,
but one can easily show that any spatial distribution can
fit the 
Log$N$-Log$S$ diagram, given some GRB luminosity function
and a cosmological model (see e.g. Rutledge et al., 1995,
Reichert and M\'esz\'aros, 1997,
M\'esz\'aros and M\'esz\'aros, 1995, Hakkila et al., 1996).
\begin{figure}[t]
\leavevmode%\centering
\centerline{
\psfig{file=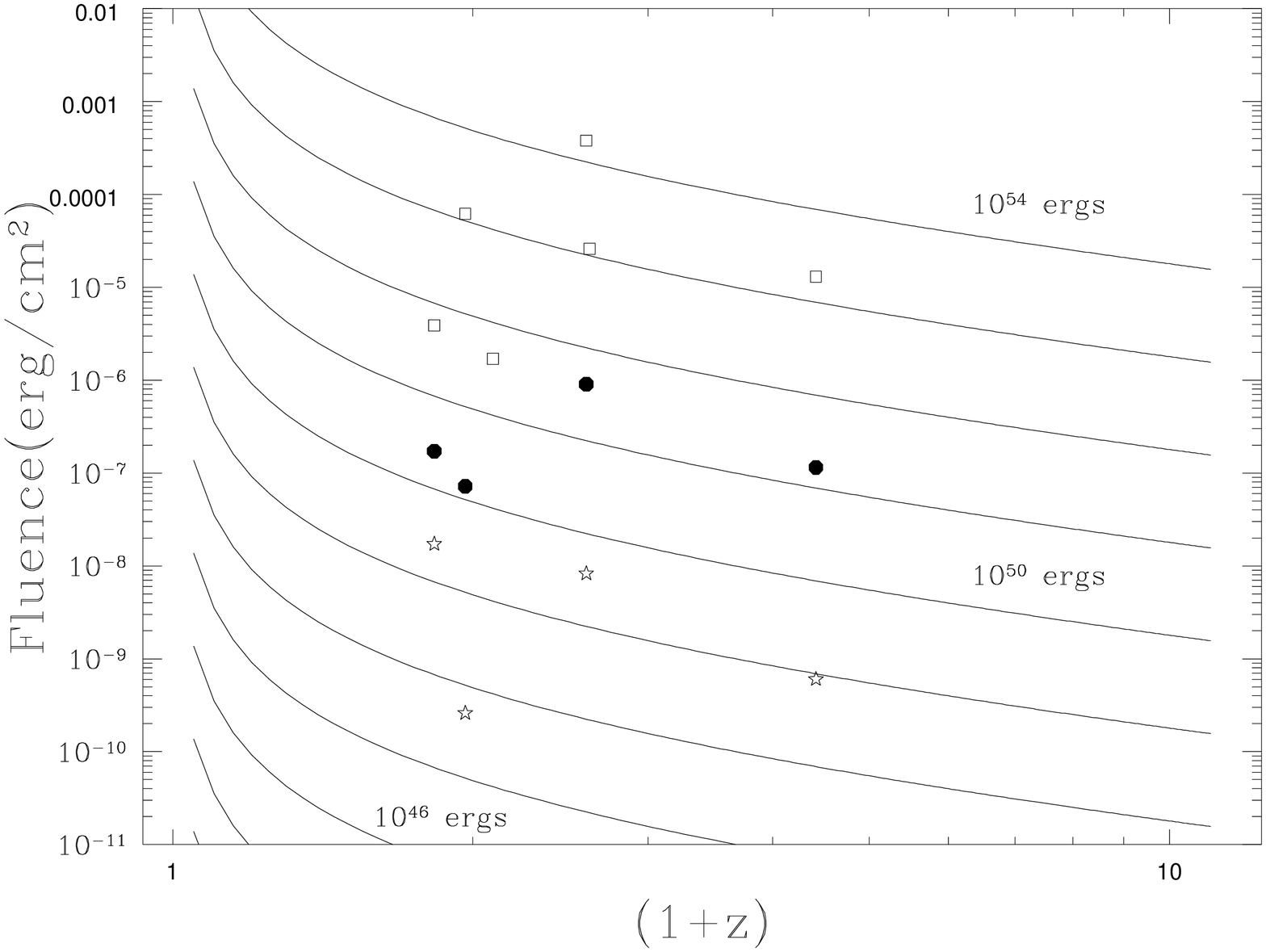,width=0.5\textwidth,height=0.5\textwidth,angle=270}
\psfig{file=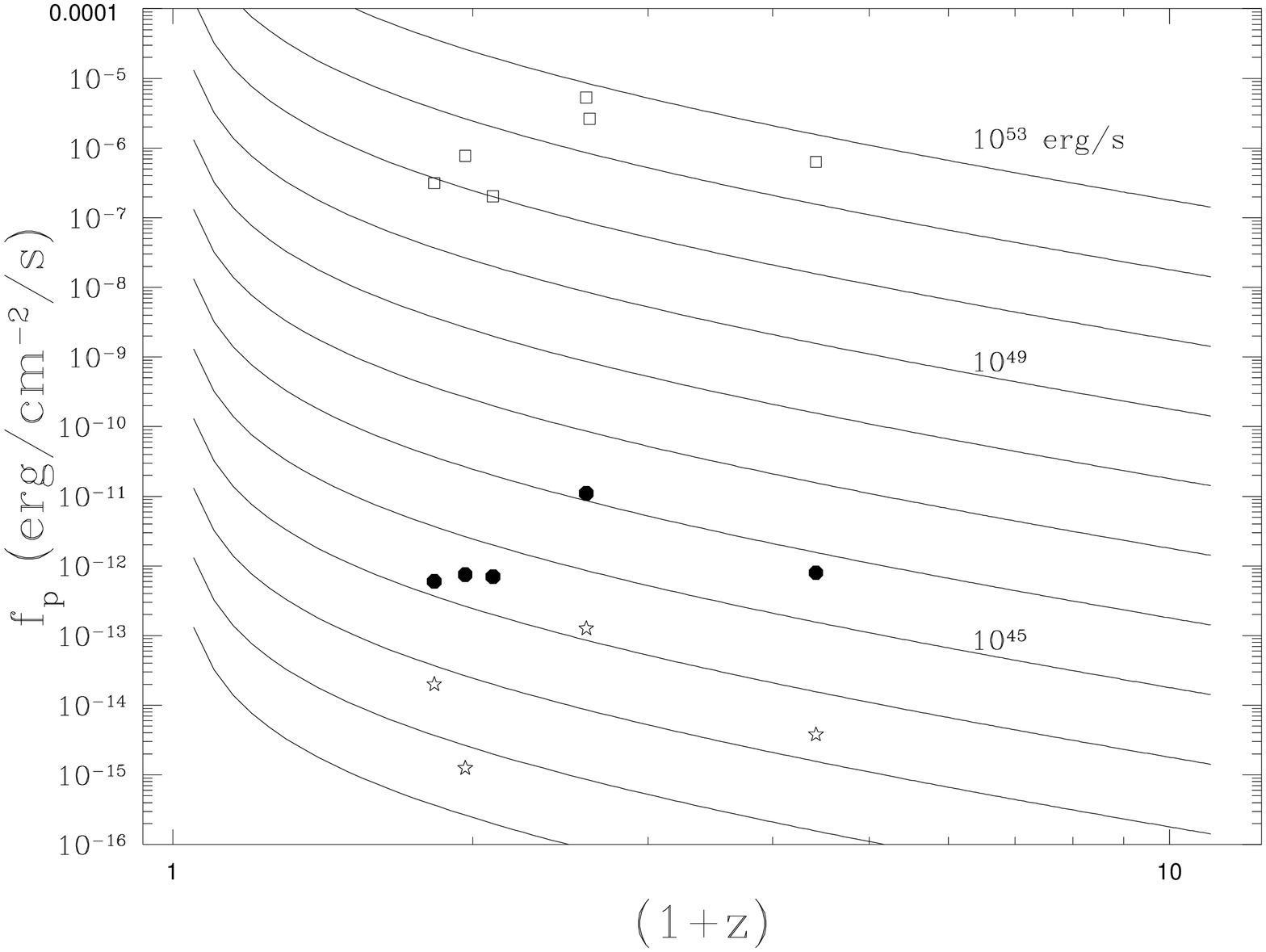,width=0.5\textwidth,height=0.5\textwidth,angle=270}
}
\caption{\, Hubble diagram for a) fluences (left panel)
and b) peak fluxes (right panel) in the
 gamma-ray 
(squares), X-ray (filled circles) and optical (stars) energy
ranges for GRBs with known 
redshifts. The solid lines are the expected relation in the 
Einstein-de Sitter cosmological model (with Hubble constant
of 60 km/(s Mpc)) for indicated total 
radiated energy in (a) or average luminosity in (b).
In Figure 1b, The X-ray fluxes are the values 8 hours after the burst, and
the optical fluxes are either the peak or the earliest detection
flux.
}
\end{figure}
  A more useful way to gain insight
into the distribution and energetics of GRBs is to study
    correlations between various burst spectral
    and temporal properties.
In this paper, we consider the
 correlation between a
characteristic photon energy, for example
the peak of the $\nu F_{\nu}$ spectrum, $E_{p}$, 
 and fluence or flux. This can 
 shed significant light on both
 the distributions and energetics of GRBs. For example, a burst
that is further away will be weaker (assuming little
luminosity evolution) and its spectrum will be redshifted;
thus one would expect a positive correlation between
a parameter such as $E_{p}$ and fluence or flux, if cosmological
effects were dominant. Intrinsic effects could either wash out
or intensify such a correlation.  Mallozzi et
al. (1995) report the presence of a 
correlation between peak photon flux and $E_{p}$,
consistent with the cosmological interpretation.
However, the degree of the correlation can be affected by
various detector
selection biases. For example, because only the brightest bursts
have adequate spectral data, many low intensity GRBs
near the detection threshold
will be missing in the sample of bursts with spectral
fits. This then will introduce
a bias  against bursts with peak counts near the threshold.
Similarly, there may exist biases against high and low
values of $E_{p}$ (Lloyd and Petrosian, 1999; hereafter LP99).
 We investigate how  the results
are modified by the truncation of the sample
due to such biases.  

 For reasons
discussed below, we focus our study in particular on total fluence, $F_{tot}$,
and $E_{p}$ correlations for a sub-sample of brighter bursts. 
Our main goal is to determine the degree of correlation between
these quantities, and investigate to what extent the correlation is due
to cosmological redshift and to what extent it is a signature of an
intrinsic relation between radiated power and spectral features.
The former signature will be useful for investigations of spatial
distributions and evolutionary processes associated with GRBs, while
the latter will be important in constraining models of energy production
and radiation processes in these sources.
We find
that there is a strong correlation between $F_{tot}$ and $E_{p}$, and
that this can be attributed to intrinsic effects, rather than to cosmological
expansion or evolution of burst properties.  We will briefly discuss
the significance of these results for particular GRB models.
   
 Getting
an accurate handle on the amount of correlation between
burst properties proves to be a challenging task, not
only because we do not have redshift measurements for most bursts,
but because we must account for the various biases and selection
effects present in the GRB data.
 Because all
detectors are sensitive over a finite energy range, and - in
the case of BATSE - are
subject to some trigger condition, selection effects are
inevitable.   For instance, as shown in Lee and
Petrosian (1996) and LP99, the fluence, flux
and $E_{p}$ all suffer from some sort of data truncation
imposed by the detector.  As explained below, the detector is most likely
to miss the high $E_{p}$ bursts with low strength; a bias against
high $E_{p}$ bursts with low fluence or flux will have the
effect of producing an apparent positive correlation in the
data. Hence, a simple correlation analysis
between raw values of $E_{p}$ and burst
strength without the consideration of the data truncation
can lead to erroneous correlations.
 This will have an important effect on
determining whether or not the correlations are intrinsic
or due to cosmological redshift; hence, understanding
the type and severity of truncation is necessary before
attempting to perform correlation studies.  
   In \S 2, we give a brief description of how we estimate
   the truncation imposed by the selection process, and determine
   how these truncations
   affect correlation studies.  Further details are presented
   in the Appendix where we describe the non-parametric
   methods we use for the task at hand.
   In \S 3, we describe
the data  we use, and in \S
4, we present our results for the correlation of $E_{p}$
with energy fluence and peak photon flux.  In \S 5, we discuss the implications
of these results in terms of cosmological and
intrinsic effects.  A summary and our conclusions
are presented in \S 6.

\section{Data Truncation}

  We
need to understand how the detector thresholds affect the data
analysis, and whether or not truncation effects are playing
a role in the correlation analysis.
  Here, we focus on the BATSE detector as an example.   BATSE 
will trigger and begin recording data on a burst if the
peak photon counts $C_{max}$ in some finite time interval
$\Delta t$ (= 64, 256, or 1024 ms) and energy range $\Delta E$
(between $E_{1}$ and $E_{2}$; usually between 50 and 300 keV) 
is greater than $C_{min}$, determined by the background in the
second brightest detector (out of eight detectors).
This suggests that the detector efficiency
is highest for longer duration
bursts (see Lee and Petrosian, 1997),
and for bursts with most of their photons in the range $\Delta E$.   
 As discussed in LP99, we can find the threshold(s) to
 any burst characteristic X measured by BATSE, utilizing the
 trigger condition. For each burst, we
 know X, $C_{max}$, and $C_{min}$.
  We can ask what is the possible range of
values of X that this burst can have and still trigger the instrument.
In general, X can have
an upper limit $u$, a lower limit $l$, or both limits; $X  \in T =
[l,u]$.  
 
 For example, we can determine
 the detection threshold of 
the observed energy fluence $F_{obs}$ (or
peak flux $f_{p}$); in this case,
there is only a lower limit $F_{obs,lim}$ (or $f_{p,lim}$).
 This threshold
is obtained from the simple relation (Lee and Petrosian, 1996)
\begin{equation}\label{equation}
f_{p}/f_{p,lim} = F_{obs}/F_{obs,lim} = C_{max}/C_{min}.
\end{equation}
The last equality is valid if
the burst's spectrum does not change drastically
throughout its duration.  
For other bursts characteristics, such a basic relation may not hold.
For example, the break energy, $E_{B}$, or 
the peak energy, $E_{p}$, of the $\nu F_{\nu}$ spectrum of GRBs
has both an upper and lower limit.  The values
of $E_{p, max}$ and $E_{p, min}$ are not related to $C_{max}/C_{min}$
and $E_{p}$ in a simple form as above, and thus we require a
more complex procedure to determine
their values.  This was discussed in detail in LP99 and
is summarized below.

To get a qualitative idea of how truncation could come into
play, consider the following:
We characterize the
  spectra of the bursts by four parameters:
  the low energy photon power 
law index, $\alpha$, the 
   high energy index, $\beta$, the break energy $E_{B}$ (or 
  $E_{p} = E_{B}(2+\alpha)/(\alpha-\beta)$ which is the peak of the $
\nu F_{\nu}$ spectrum when $\alpha > -2$ and
$\beta < -2$),
 and  the normalization factor $A$. For an example, see equation (5) 
in \S 3, which is the parameterization used by Mallozzi et al. (1999,
in preparation).    
Given a burst's
  spectrum $f_{\alpha,\beta,A}(E,E_{p},t)$, the
 observed
 fluence is $F_{obs} = \int_{0}^{T} dt
  \int_{E_{1}}^{E_{2}}E F_{\alpha,\beta,A}(E,E_{p},t) dE$
  (where $T$ is burst duration). 
 The
limiting value(s), $E_{p,lim}$('s), that $E_{p}$ can take on and still
 trigger the BATSE instrument must satisfy the equation

\begin{equation}\label{equation} F_{obs,lim} = \int_{0}^{T}dt
 \int_{E_{1}}^{E_{2}} E f_{\alpha,\beta,A}(E,E_{p,lim},t) dE.
\end{equation}

\noindent  In general, there
will be two values of $E_{p,lim}$ that satisfy
this equation.  To find these two solutions, we start with the observed
value of $E_{p}$ (for which
the right side is equal to $F_{obs}$),
and increase it (while keeping
 $\alpha$, $\beta$, and the total fluence, $F_{tot}
 = \int_{0}^{T} \int_{0}^{\infty} E f_{\alpha,\beta,A}(E,E_{p},t) dE$,
 equal to their determined values)
until the above equality is reached  - that is, until the observed
fluence is brought to the level of the threshold.  This value
is the upper limit to $E_{p}$, $E_{p,max}$.
We then decrease $E_{p}$
until this equation is again satisfied; this gives the lower
limit $E_{p,min}$.  The choice of what is kept constant
depends on the investigation at hand. 
We choose to keep the total fluence constant
because we are interested in the correlation between $E_{p}$ and
$F_{tot}$ (for other types of investigations, one might keep, say,
the normalization $A$ constant).    
 
However, in reality, BATSE triggers on counts, which is the convolution
of the impinging photon flux with the detector response matrix, DRM.
The DRM converts photons of a particular energy into detector ``counts''
over a range of energies. This allows for the possibility of 
photons from higher energies to spill over into the
``observed'' energy range.  For example, given some $C_{max}$ and $C_{min}$,
and the detector response
matrix, $DRM_{i,j}$ (where   $i$ and $j$ index the energy bins
for the incident photon flux and counts, respectively),
the correct value for $E_{p,lim}$ is obtained from the
following expressions:

\begin{equation}\label{equation}
    C_{min,j}= \sum_{i} DRM_{i,j}*
    \int_{o}^{T} f_{\alpha,\beta,A^{'}}(E_{i},E_{p,lim},t)\Delta E_{i} dt,
\end{equation}
\begin{equation}\label{equation}    
    C_{min} = \sum_{Ch2+3,j} C_{min,j}.
\end{equation}
       
\noindent where $A^{'}$ is the normalization in the spectrum appropriately
adjusted to keep the total fluence constant. In other words, we
have the incident photon spectrum characterized
by $\alpha$, $\beta$, $E_{p}$, and $A$
in $n$ energy bins of width $\Delta E$ indexed by $i$.  To get the
counts in $m$ energy bins indexed by $j$, we multiply our photon
model vector (of $i$ components) 
by the detector response matrix - an $m\times n$ matrix,
which converts photon flux into a count rate.  We then sum over the energy bins
for the count rate, 
corresponding to detector channels 2 and 3 ($\sim$ 50-300keV); this gives
us the channel 2+3 count rate on which the detector triggers.

Equation (2), on the other hand, assumes that the
DRM is strictly diagonal. Hence, the limits obtained
based on this equation are only
estimates of the true observational limits.  Including the actual
DRM would cause some spillover 
from channel 4, and contribute to the counts in channels
2 and 3.  Preliminary analysis incorporating
the DRM for a handful of bursts
shows that including the precise detector
response tends to increase the upper limit
on $E_{p}$.  In a future publication, we will
present results which show how incorporating the
actual shape of the DRM affects the limits of $E_{p}$.
In this paper, we use a sample of bright bursts, which
has a narrow
distribution of $E_p$'s and consequently requires a small correction due to
instrumental biases (e.g. 
a delta function distribution requires zero correction; see
also Brainerd, 1999).
Thus, the truncations on $E_{p}$ have little effect
on the final results of our correlation studies for the chosen
sample described in the next section.
 Nonetheless, we present
results in which the effects of the limits on fluence
and $E_{p}$ are taken into account.  
 In the
Appendix, we present the results on how to perform correlation studies
on data which suffer from truncation; these are based on new non-parametric
techniques developed by Efron and Petrosian (1999). 
Once the truncation of the data is
determined, we can then
proceed to the investigation of correlations.
  
\section{Data}
    Our sample consisted of a set of bursts from the 4B catalog (Paciesas et al., 
    1999)
    which have 16 channel CONT (continuous) data.  Using these
     data, the bursts are fit
    to a Band Spectrum (1993).  The spectrum
    is parameterized as follows:

\begin{equation}
  f(E) =
\cases{ A (E/E_{B})^{\alpha} {\rm exp}[-(2+\alpha)(E/E_{p})],
&
  $ E < E_{B}$ \cr
A (E/E_{B})^{\beta}{\rm exp}[\beta-\alpha],&
$E >   E_{B}$}
\end{equation}

%Mallozzi's way:
%\begin{equation}
%  F(E) =
%\cases{ A (E/100keV)^{\alpha} {\rm exp}[-(2+\alpha)(E/E_{p})],
%&
%  $ E < E_{B}$ \cr
%A (E/100keV)^{\beta}{\rm exp}[\beta-\alpha] [(\alpha-\beta)E_{p}/
%(100keV(2+\alpha))]^{\alpha-\beta},&
%$E >   E_{B}$}
%\end{equation}

\noindent 
where $A$ is units of $\rm ph/cm^2/s/keV$. Note that for $\beta < -2$
and $\alpha > -2$, $E_{p}$ is
the maximum of $\nu F_{\nu} \propto
E^{2}f(E)$. Otherwise,
$\nu F_{\nu}$ has no maximum
and $E_{p}$ is some characteristic photon energy.
%(in
%the case of $\beta > -2$
%and $\alpha < -2$, $E_{p}$ is the {\em minimum} of $\nu F_{\nu}$.
The parameter
$\alpha$ is always greater than $-2$ in
the fitting process so that $\nu F_{\nu}$ always converges
for small $\nu$; but $\beta$ may exceed $-2$ which implies the peak
of $\nu F_{\nu}$ must occur outside the range of detector sensitivity.
The LP99 sample - based on four channel LAD data - contains a
complete sample of bursts with known $C_{max}$ and $C_{min}$ values and 
therefore
has a well defined truncation. Unfortunately the present 16 channel
sample (containing a total of 653 bursts
with acceptable fits for the fluence (time averaged) fits, and 655 bursts
with acceptable fits for the peak flux fits) -  
although
more reliable in the values of the spectral parameters - does not have a well
defined selection criterion.  Because the brightest bursts give the best fits,
%this sample includes bright bursts with no known values of $C_{min}$; 
%furthermore,
most bursts in this sample 
with a known $C_{min}$ value 
have $C_{max}/C_{min} \gg 1$. These will obviously
  suffer less truncation effects than bursts with counts
  closer to the threshold; bursts with 
$C_{max}$ near $C_{lim}$ will have one 
or both $E_{p,lim}$'s near $E_{p}$ and the truncation
will be an important effect.
In other words, the sample of bursts
with reliable spectral fits is not complete to within
well known limits, and the
selection process most probably has introduced complicated truncations.
Figures 2a and 2b show the sample of bursts which had spectral
fits (circles),
and those bursts in the catalog with known $C_{max}$ and
$C_{min}$ values but no spectral fits (squares),
 in the $F_{tot}-F_{tot,lim}$  and $f_{p}-f_{p,lim}$ planes,
 respectively.  
\begin{figure}[t]
\leavevmode%\centering
\centerline{
\psfig{file=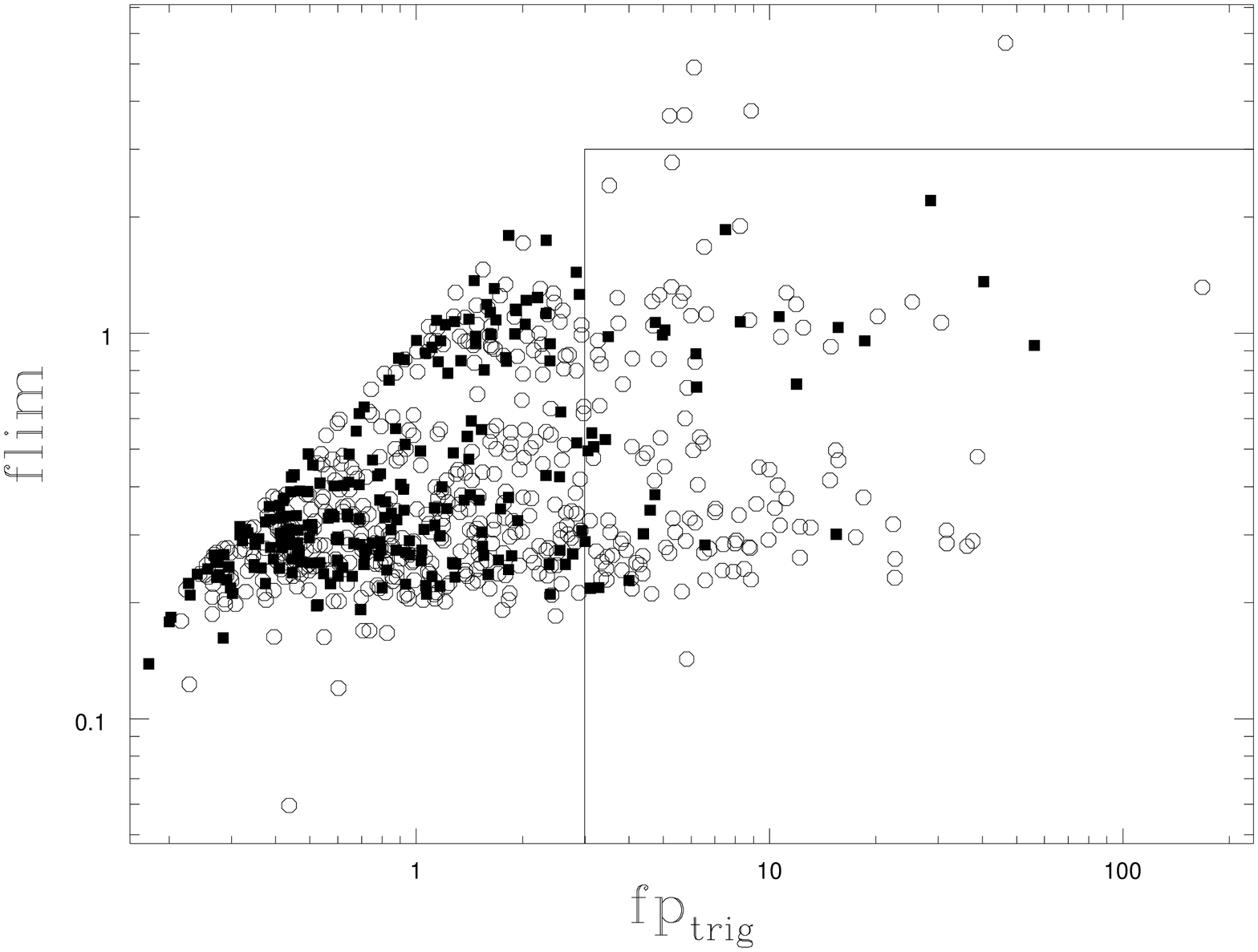,width=0.5\textwidth,height=0.5\textwidth,angle=270}
\psfig{file=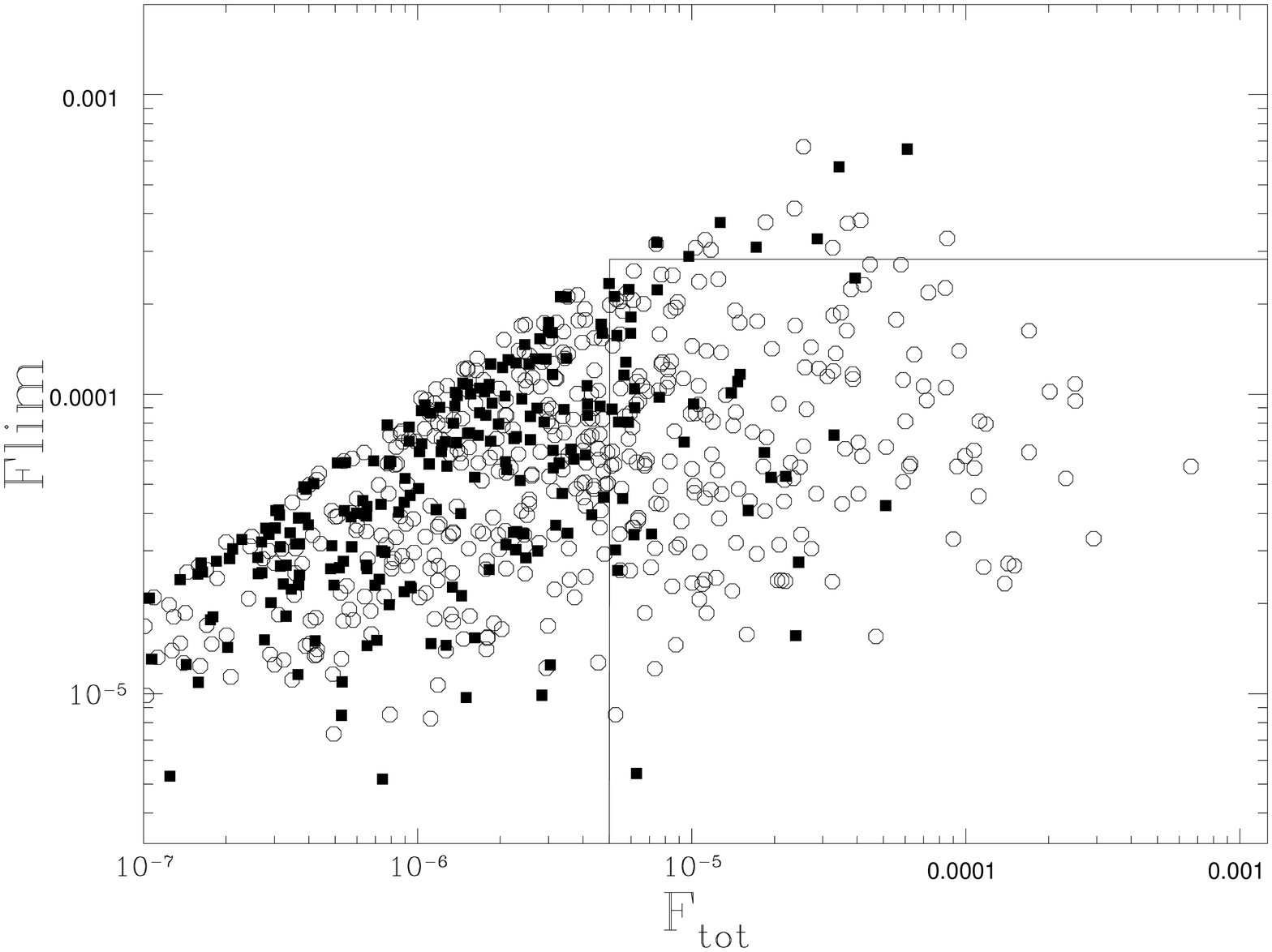,width=0.5\textwidth,height=0.5\textwidth,angle=270}
}
\caption{\, a) Peak flux, $f_{p}$ vs. limiting flux, $f_{lim}$ (left panel)
and
b) total fluence, $F_{tot}$ vs. limiting fluence $F_{lim}$ (right panel).
The circles indicate the bursts with spectral fits, and the black squares
mark those bursts in the catalog which do not have spectral fits.  The
box indicates our ``truncated'' sample, in which we attempt to get
the most complete sample of bursts with spectral fits (i.e. fewest black
squares in our sample), while still keeping a reasonable number of bursts
for our statistical analysis.
}
\end{figure}
As evident, there is not
 a clear indication of the nature of the selection
 process. Without exact knowledge of this selection
criteria, we cannot account for data truncations properly or determine 
correlations
accurately.  
 To circumvent this situation, we choose the following sub-samples
 which have a better
defined selection criterion.  We truncate the available data 
parallel to the axes in Figure 2a and 2b, so that 
the
above mentioned uncertainty is minimized.
The truncation chosen here is a compromise
which involves trying to get as large a sample
(open circles) as possible, without too many missing
bursts (black squares).  That is, we attempt to
minimize the bias without reducing the sample
to an unreasonably small size.  There are still a few bursts
without spectral fits in this ``complete'' sample; however, we
assume that these few bursts will not significantly alter
the final result.
For peak flux, the truncation was
made at $f_{p} \ge 3.0$ ph/(cm$^{2}$ s), for  the observed fluence (in the 
range
50-300 keV) we select a subsample with $F_{obs}\ge 10^{-6}$ ergs/cm$^{2}$, and
for the summed fluence (summed over all four LAD channels  
($~20 \rm keV- \sim 1.5 \rm MeV$)) we
select sources with $F_{sum} \ge 5 \times 10^{-6}$ ergs/cm$^{2}$.

\section{Results}
 The results of our investigation of the 
 correlation between peak flux 
and fluences 
with $E_{p}$ are summarized in 
Figures 3a and 3b, and in Table 1.  Note that $F_{sum}$ is
approximately equal to the total fluence of the burst, 
$F_{tot}=\int F(E)dE$ for $E_{p}$'s in the
energy
range that we are dealing with.  
The
fluences are correlated with the burst {\em average} $E_{p}$, while the peak
photon flux is correlated with the value of $E_{p}$ at the time of the peak.
Figures 3a and 3b show binned values of ${E_{p}}$ versus $f_{p}$ and
$F_{tot}$, respectively, for the whole sample without any consideration
of completeness or detector bias (dotted histograms), as well as
our more limited but complete sub-samples (solid histograms).
 The ${E_{p}}$ versus $f_{p}$ behavior for the
whole sample is
in agreement with the results reported by Mallozzi et al. (1998) on
a slightly different sample .  While there clearly is a
significant correlation (particularly in the low end of the
$f_{p}$ distribution), it is not
known how much of this is due to incompleteness of the
data or other selection biases.  The more complete sub-sample (solid
histogram) shows
very little correlation.  The Kendell's $\tau$ test (see Appendix
for further discussion) results shown
in Table 1 are in agreement with the impressions from Figure 3a and
quantify the differences between the whole sample and the sub-sample.  The
column labeled 
{\em raw result} shows the significance of the correlation, $\tau$,
without accounting for truncation of the variables, while the 
{\em corrected result}
uses the techniques described in the Appendix
 to account for these truncations.
While the total sample shows $\tau$ values of $\sim 8\sigma$
for the correlation with $f_{p}$, the complete
sample valid over a limited range of peak flux shows no significant
($<3\sigma$) correlation.  The correction for data truncations
alter the results only slightly because of the narrowness of
the $E_{p}$ distribution, as mentioned above.
It should be noted that the lever arm is small and most of the $E_{p}$
- peak flux correlation {\em may} be contained in the low peak flux
bursts; again, however, this is where the selection bias is the strongest 
so we cannot be certain that this is a real correlation between
$E_{p}$ and $f_{p}$.

\begin{center}
\centerline{\underline{TABLE 1}}
\centerline{Simulation results for the case when truncation eliminates the
correlation.}
\begin{tabular} {lcccr} \hline \hline
   Correlation           & raw result   & corrected result \\ \hline
$F_{obs}$ w/ $E_{p}$ (whole) & (+) 5.1 $\sigma$ & (+)8.0$\sigma$
\\ 
$F_{obs}$ w/ $E_{p}$ (sub-smp) & (+)  5.6 $\sigma$ & (+) 5.5$\sigma$
\\ 
$F_{sum}$ w/ $E_{p}$ (whole) &(+) 7.5$\sigma$  &
 (+)9.4$\sigma$
\\ 
$F_{sum}$ w/ $E_{p}$ (sub-smp) & (+)  6.5$\sigma$   &
 (+)5.8$\sigma$
\\ 
$f_{p.trig}$ w/ $E_{p}$ (whole) &(+) 7.9$\sigma$ &
(+)9.3$\sigma$
\\
$f_{p.trig}$ w/ $E_{p}$ (sub-smp) & (+) 2.5$\sigma$ &
(+)2.6$\sigma$
\\ \hline \hline
\end{tabular}
\end{center}

 The behavior of the correlation between fluences and $E_{p}$
is quite different. Although the significance of the
correlation is large for the sample as a whole,  Figure 3b shows
that the functional dependence of the correlation is much stronger for
high (solid histogram)
rather than low valaues of the fluence. Again, because of the
presence of many  bursts without spectral fits in the whole sample, we
do not know how accurate the functional form of the correlation
in the low end of the fluence distribution is.
In what follows, we shall use the truncated sub-sample
of Figure 2b to search
for cosmological signatures, because it shows a significant correlation
and is more robust, due to our well defined truncation parallel
to the axes in the $F_{tot}-F_{tot,lim}$ plane. 

\begin{figure}[t]
\label{3}
\centerline{
\psfig{file=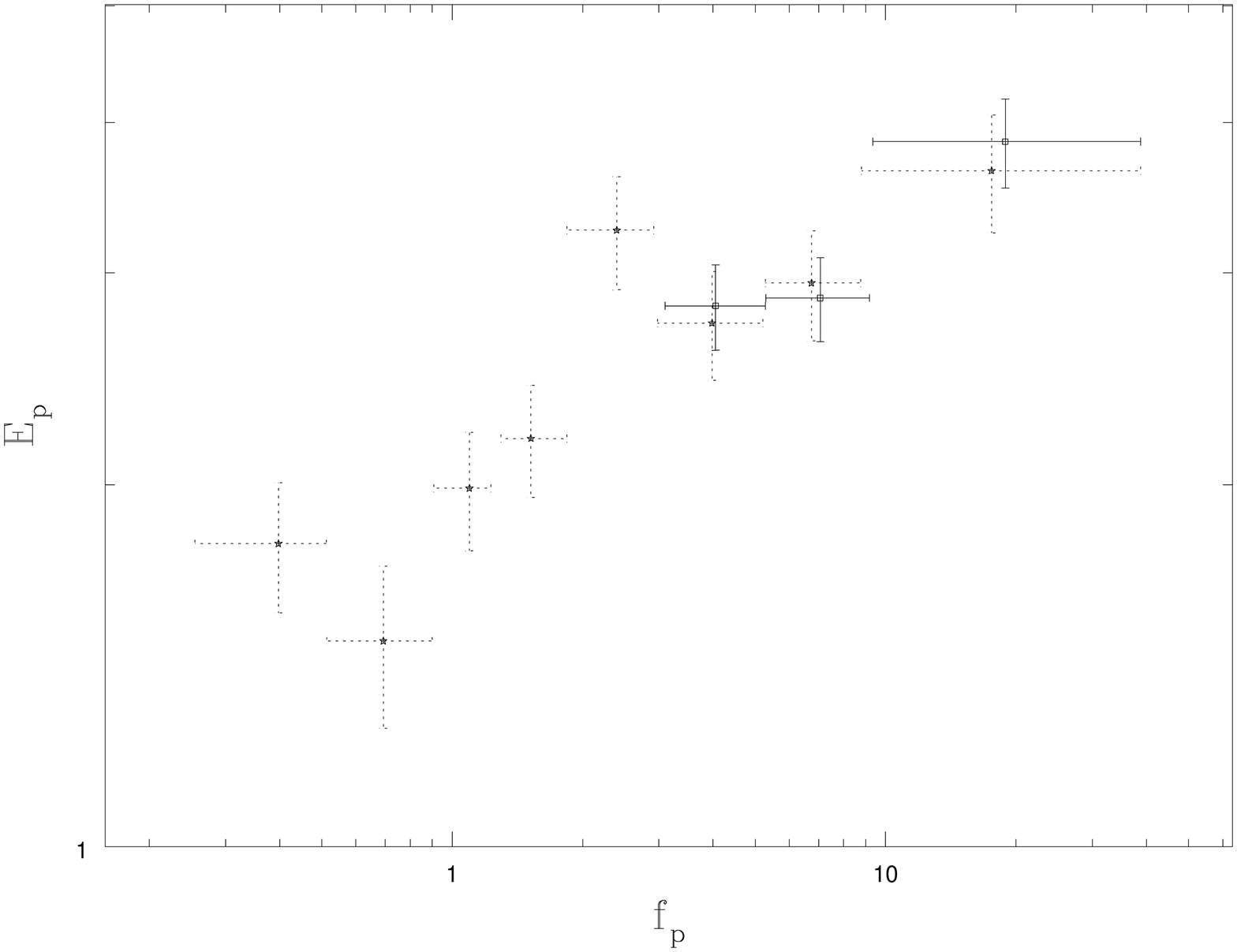,width=0.5\textwidth,height=0.5\textwidth,angle=270}
\psfig{file=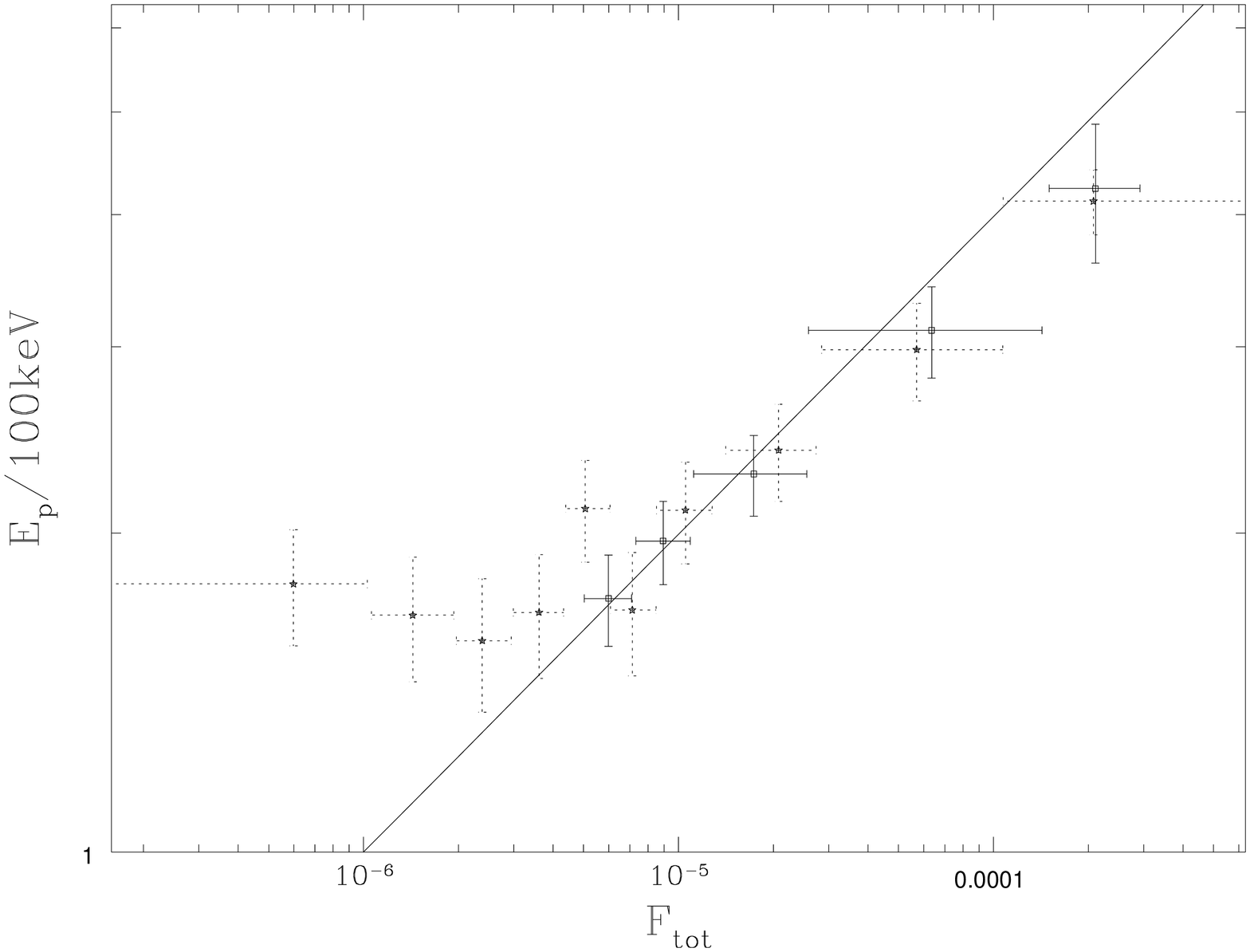,width=0.5\textwidth,height=0.5\textwidth,angle=270}
}
\caption{
Average peak energy, ${\overline{E_{p}}}$ vs.
 a) peak flux, $f_{p}$ (left panel) or b) total
fluence, $F_{tot}$ (right panel)
for the whole (dashed histogram) and sub (solid histogram) spectral sample.
Notice the flux shows a strong correlation at low values, but much less
of a correlation for the brighter bursts.
In the fluence case, 
the more complete sub-sample carries most of the correlation.  The
solid line is a least squares fit to the data which is in 
agreement with the functional form of the correlation we obtain from our
statistical methods.}
\end{figure}

   A least squares fit to log$({E_{p}})$ - log$(F_{tot})$
   of the sub-sample (solid histogram) in Figure 3b gives
   a slope of $0.28 \pm 0.04$ (solid line).
   Our statistical methods presented in the Appendix give a similiar
   result: We define
   a new variable $E_{p}^{'} = E_{p}F_{tot}^{-\delta}$ and
   calculate the Kendell's $\tau$ statistic for the new variable $E_{p}^{'}$
   and $F_{tot}$ as a function of the parameter $\delta$.  Figure 4 shows
   this variation.  The range of $\delta$ such that
    $|\tau| < 1$ determines the correct functional form
   form of the correlation.  We find $\delta=0.29\pm 0.03$, meaning
   $E_{p} \propto F_{tot}^{0.29}$, in good agreement with the above
   least squares fit result.

\pagebreak   

\section{The Nature of the Observed Correlation}
  
\subsection{The Cosmological Correlation}

A positive correlation between 
an observed burst strength ($f_{p}$ or $F$)
 and $E_p$ follows the trend  expected 
from cosmological effects; bursts which are further
away have lower intensity as well as a lower (redshifted)
value of $E_{p}$.  We now test to see if the observed 
correlations reported above
 can be attributed {\em fully} to these 
effects.  We will focus particularly on the $F_{sum} \approx
F_{tot}$ results, because - as mentioned above -
it is our most robust result, and
because the total fluence can be related to the total
radiated energy 
and the redshift of the burst, without any need for the
so-called K-correction or correction
for duration bias (Lee and Petrosian, 1997);
 $F_{tot}={\cal E}_{ \rm rad}/(\Omega_{b} [d_{\cal{E}} (\Omega_i,z)]^{2})$, 
where ${\cal E}_{ \rm rad}$ is  the total radiant 
energy  (in the gamma-ray range),  $\Omega_b$ is
 the average beaming solid angle, $\Omega_i$ 
denotes the cosmological 
model parameters, and $d_{\cal{E}} (\Omega_i,z) =
d_L(\Omega_i,z)/\sqrt (1+z)$ with $d_L$ as the usual bolometric luminosity 
distance.  We also define $E_{po}$ as the value
of the peak of the $\nu F_{\nu}$ spectrum
in the rest frame of the burst, so that 
$E_{p} = E_{po}/(1+z)$. 
For the task at hand, we need to specify  a cosmological model, as well as the 
distribution function of redshift and intrinsic burst parameters,
$\Psi(E_{p},\Omega_b,{\cal E}_{ \rm rad},z)$. Note that because 
the beaming angle enters always in 
conjunction with ${\cal E}_{ \rm rad}$,  we assume 
a delta function distribution of $\Omega_b$ so that it can be eliminated from 
the distribution function. This amounts to replacing ${\cal E}_{ \rm rad}$ with  
${\cal E}_{ \rm rad}/\Omega_b$. 
We also assume 
that the intrinsic parameters ${\cal E}_{ \rm rad}$ and $E_{po}$
do
not evolve with
cosmic time and are not correlated.
These assumptions 
mean that the multivariate distribution function becomes separable;

\begin{equation} 
\Psi(E_{po},\Omega_b,{\cal E}_{ \rm rad},z) = 
\phi({\cal E}_{ \rm rad})\zeta(E_{po})\rho(z).
\end{equation}
 
\noindent If now we change variables from ${\cal E}_{ \rm rad}$ to
$F_{tot}$ and $E_{po}$ to $E_{p}$, we get 
the joint distribution of observed $E_p$ and $F_{tot}$ as

\begin{equation}
	{d^2N(E_{p}, F_{tot}) \over dE_pdF_{tot}} =  \int_{0}^{\infty} dz 
(dV/dz) \rho (z) [d_{\cal{E}} (\Omega_i,z)]^{2}
 \phi(F_{tot}[d_{\cal{E}} (\Omega_i,z)]^{2})
(1+z)\zeta(E_{p}(1+z)),
\end{equation}

\noindent where $dV/dz$ is the differential of the co-moving volume up to $z$.
From this we can compute the individual distributions, and
average value of $E_{p}$ as a function of $F_{tot}$. For 
example,   

\begin{equation}
	\overline{E_p(F_{tot})} = \frac{\int dE_p E_p [d^2N(E_{p}, 			
         F_{tot})/dE_pdF_{tot}]}{dN(F_{tot})/dF_{tot}},
\end{equation} 

\noindent where the differential source count in terms of fluence is

\begin{equation}\label{Fdist}
	dN(F_{tot})/dF_{tot} =\int dE_p [d^2N(E_{p},F_{tot})/dE_pdF_{tot}].
\end{equation}

\noindent [In an analogous manner, we can find the differential
distribution of $E_{p}$ and the average value $\overline{F_{tot}(E_{p})}$.]
We now have an expression for average observed $E_{p}$ as a function
of the observed total fluence.  Implicit in this expression is the
cosmological contribution to the correlation between $\overline{E_p}$ and
$F_{tot}$.  To see if this is the form of the correlation observed in
the data, we carry out the following tests:
We assume various plausible models for the functions $\zeta$, $\phi$, 
and $\rho$, and
compute the expected cosmological relation $\overline{E_p(F_{tot})}$. We then
remove this correlation from the data by the transformation
$E_{p}^{'} = E_{p}/\overline{E_p(F_{tot})}$,  and see if we
are left with any correlation between the observed $F_{tot}$ and $E_{p}^{'}$ 
distributions. 
Only if none remains (i.e. $|\tau| < 1$, where - again - $\tau$
indicates the significance of the correlation between $F_{tot}$ and
$E_{p}$),
then we can attribute the correlation between $F_{tot}$ 
and $E_{p}$
to cosmological effects alone.

Assuming no intrinsic
correlation between ${\cal E}_{ \rm rad}$ and
$E_{po}$ as in equation
(6), we try several models for the functions $\phi$, $\zeta$, and $\rho$.
We assume that $\zeta$ is a Gaussian  with a mean value for $E_{po}$ 
of $Q$ and a dispersion $\sigma_Q$, $\phi$ is either a delta
function (standard candle) or a power law 
 in the radiated
energy,   and $\rho$ is either a constant or follows
the star formation rate.
We present results for the Einstein de-Sitter 
cosmological model, and adopt a Hubble constant
of $60 \rm \ km \ s^{-1} Mpc^{-1}$; however, we find the quantitative results
are insensitive to the cosmological model we choose.  For example, a vacuum
dominated universe with $\Omega_{\rm matter} = 0$ and $\Omega_{\Lambda} = 1$ 
(where
$\Lambda$ is the cosmological constant),
leads us to almost exactly the same results as
a matter dominated Einstein-de Sitter universe with
 $\Omega_{\rm matter} = 1$, $\Omega_{\Lambda} = 0$.  

Our general conclusion is that, for the
more plausible values of the parameters for 
this
set of distributions, most
of the observed correlation cannot be attributed to cosmological
effects alone, and must be intrinsic to the emission mechanism
or produced by additional evolution of the sources.  
We give two examples below:

\begin{figure}[t]
\leavevmode%\centering
\centerline{
\psfig{file=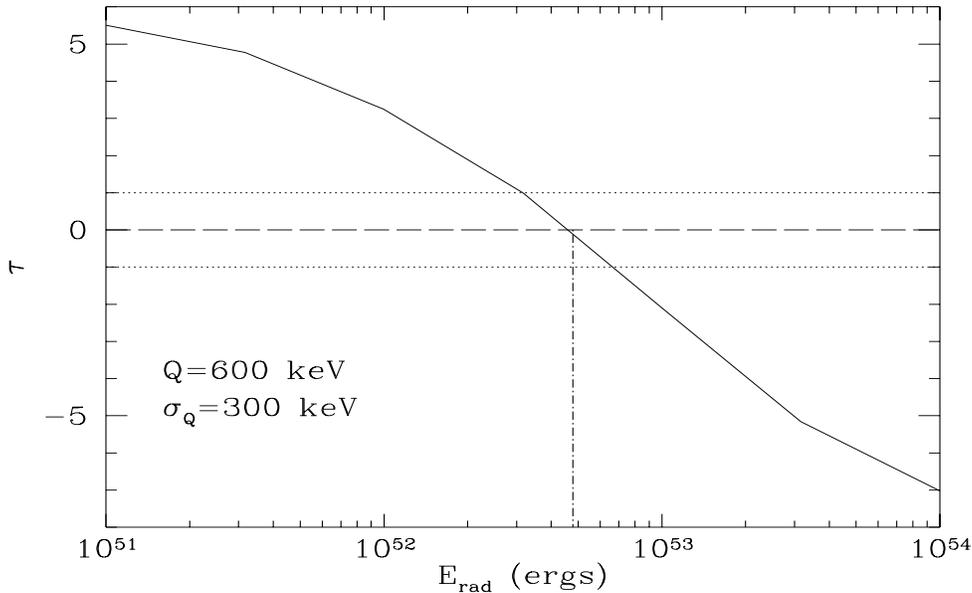,width=0.8\textwidth,height=0.5\textwidth,angle=270}
}
\caption{\,  Significance of the correlation, $\tau$, vs. radiated
energy, ${\cal{E}}_{\rm rad}$, for the model given in case 1 of \S 5.1
in the text.
Here, we find that the correlation is removed for a standard candle
energy of $5\times 10^{52}$ erg.  This result is independent of
the GRB density distribution.  
}
\end{figure}
1) For a {\em standard candle radiated energy}, $\cal{E}_{\rm *}$, i.e.
 $\phi = \delta({\cal E}_{ \rm rad}-\cal{E}_{\rm *})$, 
 our expression for the average
$E_{p}$ as a function of total fluence becomes: 
\begin{equation}\label{equation}
  \overline{E_p(F_{tot})} = \frac{\int dE_{p} E_{p} {\rm exp}[-(E_{p}Z_{o}-Q)^{2}
  \sigma_{Q}^{2}]}
  {\int dE_{p} {\rm exp}[-(E_{p}Z_{o}-Q)^{2}/\sigma_{Q}^{2}]}
\end{equation}
\noindent where $Z_{o}^{1/2} =
1+ ((H_{o}/2c)^{2}{\cal{E}}_{\rm *}/F_{tot})^{1/2}$.  
Note that because
of the delta function in energy, we are able to eliminate the redshift
integral (a single redshift is picked out),
and the expression becomes independent of the rate density
evolution.  In Figure 4,
we plot $\tau$ as a function of $\cal{E}_{\rm *}$.  Here we see that the correlation
{\em can} be removed ($|\tau| < 1$) for $\cal{E}_{\rm *} \rm = (5 \pm 2)
 \times 10^{52}$ ergs (this
result is fairly insensitive to the value of $Q$ and $\sigma_{Q}$ of the
$E_{po}$ distribution).   However, we know from the bursts with
measured redshifts (e.g., see Figure 1) that a delta function in burst
radiated energy is not a good approximation.
  Hence we try a more realistic model
for the distribution of burst parameters.

2) Here, we choose a {\em power law distribution
in the radiated energy}, $\phi={\cal E}_{ \rm rad}^{-\delta}$,
with upper and lower
cutoff $\cal{E}_{\rm max}$ and $\cal{E}_{\rm min}$.  Again, the
expression for $\overline{E_p(F_{tot})}$ is obtained from equations (8)
and (9) with
 
\begin{equation}\label{equation}
d^2N(E_{p},F_{tot})/dE_pdF_{tot} = \int_{z_{min}}^{z^{max}}
dz\frac{dV}{dz} (1+z) [d_{\cal{E}} (\Omega_i,z)]^{2-2\beta} \rho(z) 
{\rm exp}[-(E_{p}(1+z)-Q)^{2}
  \sigma_{Q}^{2}]
\end{equation}

%\overline{E_p(F_{tot})} = \frac{\int dE_{p} E_{p} \int_{z_{min}}^{z^{max}}
%dz\frac{dV}{dz} (1+z) [d_{\cal{E}} (\Omega_i,z)]^{2-2\beta} \rho(z) 
%{\rm exp}[-(E_{p}(1+z)-Q)^{2}
%  \sigma_{Q}^{2}]}
%  {\int dE_{p}  \int_{z_{min}}^{z^{max}}
%dz\frac{dV}{dz} (1+z) [d_{\cal{E}} (\Omega_i,z)]^{2-2\beta} \rho(z) 
%{\rm exp}[-(E_{p}(1+z)-Q)^{2}
%  \sigma_{Q}^{2}]}
%\end{equation}
  
\noindent where $(1+z_{min})^{1/2} = 1+ ((H_{o}/2c)^{2}{\cal{E}}_{\rm min}/
F_{tot})^{1/2}$,
and $(1+z_{max})^{1/2} = 1+ ((H_{o}/2c)^{2}{\cal{E}}_{\rm max}/
F_{tot})^{1/2}$.  We test both 
$\rho(z)$ following the
star formation rate and $\rho(z)$ as a constant.  We use two different
 star formation rates (SFRs). In the first, based on results from 
Madau et al. (1998), the SFR peaks at $z \sim 2$,  and then
decreases slowly for high $z$.  In the second model, based on recent
results from SCUBA (Hughes et al., 1998, Sanders, 
1999), the SFR remains approximately constant for
$z > 2$.
\begin{figure}[t]
\leavevmode%\centering
\centerline{
\psfig{file=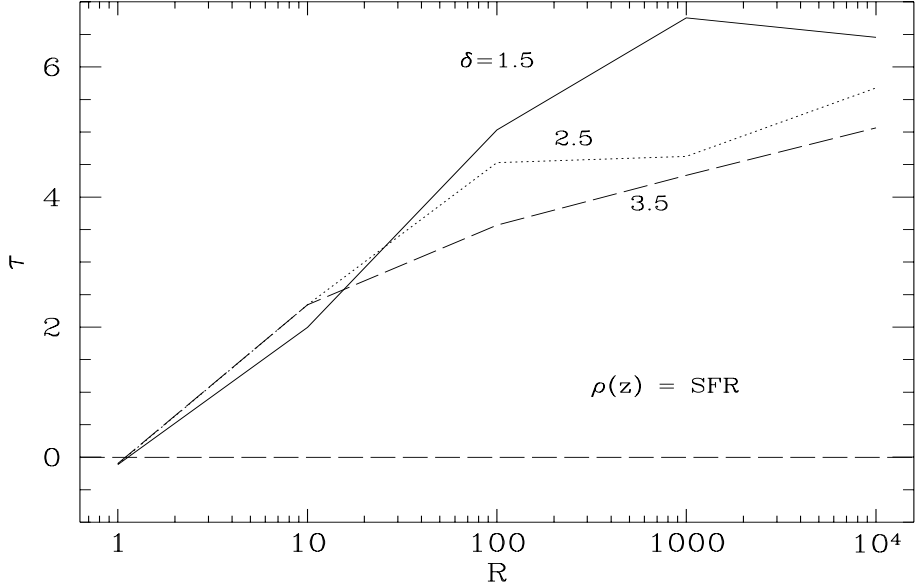,width=0.8\textwidth,height=0.5\textwidth,angle=270}
}
\caption{  Significance of the correlation, $\tau$, vs. $R \equiv
{\cal E}_{\rm max}/
{\cal E}_{\rm min}$, for the model in case 2 of \S 5.1 in the text.
The variables ${\cal E}_{\rm max}$ and ${\cal E}_{\rm min}$ are the upper 
and lower cutoffs to the burst power law luminosity
function, respectively.  Here, we have chosen $\sqrt{{\cal E}_{\rm max}
{\cal E}_{\rm min}} = 
5\times 10^{52}$, so that when $R \rightarrow 1$, we recover
our result in Figure 4 above.  The curves are shown for three different
values for the luminosity function index $\beta$, and for a density
following the star formation rate.  The results are qualitatively
similiar for a constant GRB density, or the intermediate SFR given
in the text. 
}
\end{figure}
[Note this latter
 parameterization of the SFR falls in between the Madau
parameterization and the constant rate case.]  Our results
from all three models for the co-moving rate density
are quantitatively similar (note in
the following figures, ``SFR'' will refer to the Madau parameterization).
Figure 5 plots $\tau$ vs. $R \equiv \cal{E}_{\rm max}/\cal{E}_{\rm min}$ 
for a density following
the SFR.
Here we have chosen $\sqrt{\cal{E}_{\rm min}\cal{E}_{\rm max}} = 5\times
10^{52}$ erg; note that when 
$\cal{E}_{\rm max}/\cal{E}_{\rm min} \rm = 1$ or if $\delta \gg 1$,
 we get back the delta function
case above, where the correlation is removed.  In general, for a broader,
more realistic ``luminosity function'', the correlation
due cosmological expansion is smeared out and becomes too
weak to explain the strong observed correlation.  We find a
similar result for the case when $\rho$ = constant or the intermediate
SFR.

\begin{figure}[t]
\leavevmode%\centering
\centerline{
\psfig{file=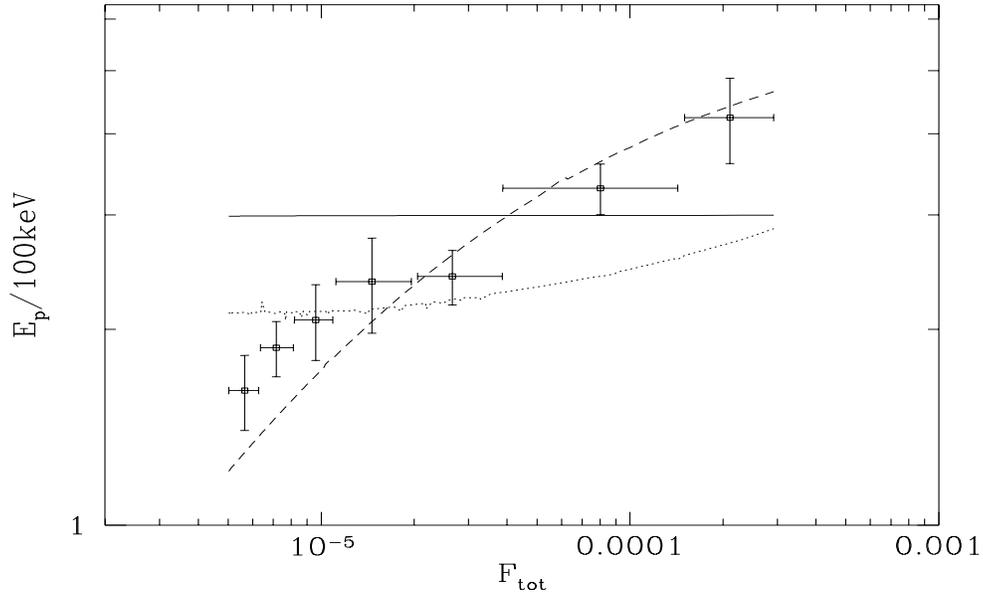,width=0.8\textwidth,height=0.5\textwidth,angle=270}
}
\label{6}
\caption{ Total fluence, $F_{tot}$,  vs. average peak energy,
$E_{p}$.  The crosses mark the data points binned (every 20) in fluence.
The curves correspond to the theoretical functions computed in
examples 1 and 2 in \S 5.1 of the text.  The dashed curve is the case of
 a constant radiated energy, ${\cal E}_{*}$.
The dotted curve is for a power law luminosity function,
with $R = 10^{4}$, and a constant GRB density.
The solid curve is the same, but with GRBs following
the Madau star formation rate.  Notice the latter two functions cannot
account for the observed correlation as shown in Figure 5 above.
The dashed curve shows the model in case 1 (standard
candle energy) can account for
the correlation, as we see in Figure 4.}
\end{figure}
Another way to view these results is shown in Figure 6.  Here, we have
plotted the theoretical curves $\overline{E_{p}(F_{tot})}$ on top of the
data for three of the cases mentioned above.  The curve from case 1 (in
which the luminosity function is a delta function (dashed curve)) 
follows the data fairly well, while the curves from the second, more
realistic cases (in which the luminosity function is a power law
and the density is either constant (dotted curve)
 or follows the SFR (solid curve)) are too
flat; the cosmological correlation is washed out by the dispersion
in the radiated energy.  The natural interpretation in 
this case is that the observed 
correlation between $F_{tot}$ and
$E_{p}$ is either caused by the cosmological evolution 
of one or more of the parameters or is due to an
 intrinsic correlation between $E_{po}$ and $\cal{E}_{\rm
rad}$.

\subsection{Evolutionary Effects}
If $\cal{E}_{\rm rad}$ and/or $E_{po}$ depend 
on redshift, we can not separate our variables as we have done
in equation (6).  In that case, we expect a different $E_{p}-F_{tot}$
correlation than in
the case when the intrinsic
variables are separable from redshift
(not evolving).  For example, if the luminosity
function evolved such that a characteristic $\cal{E}_{\rm rad}$
{\em decreased (increased)} with increasing redshift,
 then we would see a {\em stronger (weaker)} correlation due
to cosmological effects than presented in the examples given above.
This type of evolution must be explained by models of
the progenitors of GRBs, and may be difficult to achieve.  Likewise,
if the bursts' intrinsic peak energies, $E_{po}$, were on average
{\em lower (higher)} in the past, then we would see a {\em stronger
(weaker)} correlation between $E_{p}$ and $F_{tot}$ than if there were
no evolution of $E_{po}$.  We might expect some evolution
if $E_{po}$ depends strongly on the conditions of the circumburst
medium (e.g., in the external shock model discussed below), {\em and} if
the
medium itself evolved strongly with redshift.  To account for the
strong correlation we see in the data, the circumburst medium would have to
become more dense with decreasing redshift.  However,
we might expect that the progenitors
of GRBs have approximately the same type of environmental conditions; for 
example, if bursts are a result of the death of a very massive star and
occur in star forming regions as these models predict,
then we expect the circumburst densities to be about same.  However, we
know too little about the conditions of the environment around the
progenitor of GRBs to say anything definite about this possibility.

\subsection{Intrinsic Correlation} Perhaps
a more likely scenario is that the intrinsic variables
$\cal{E}_{\rm rad}$ and $E_{po}$ are correlated.  For instance, the
physics of the burst emission may lead to a case in which
bursts with more total energy radiate at higher peak energies.

To examine this possibility more closely,
we repeat the procedure described in \S 5.1 assuming a
power law parametric form for the 
intrinsic correlation between $Q$, the mean of the
$E_{po}$ distribution, and
${\cal E}_{ \rm rad}$; $Q \propto {\cal E}_{\rm rad}^{\eta}$.
We vary $\eta$ until the correlation between $E_{p}$ and $F_{tot}$
is removed ($|\tau|<1$).   For a model as in case 2
where 
$\phi = {\cal E}_{ \rm rad}^{-\delta}$, 
$\cal{E}_{\rm min}/\cal{E}_{\rm max}\rm = 10^{2.5}$ 
(a somewhat broad luminosity function, within the range of
the dispersion shown in Figure  1b),
$\delta=2.5$, and $\rho=$ SFR, we find that 
we can remove the observed correlation if 
$Q \propto {\cal E}_{ \rm rad}^{0.47 \pm 0.08}$. 
 Similarly, in the same model except with a {\em constant}
density, we find that we can remove the correlation in the data
if $Q \sim {\cal E}_{ \rm rad}^{0.62 \pm 0.1}$.  [Clearly an
intermediate exponent is expected for the SFR model that is
constant for $z > 2$.]
These results can be used to test models of GRB emission, most
of which predict some sort of relation between these quantities.

For example, if the emission can be explained by optically
thin synchrotron radiation 
with a power law
  distribution of electrons, then $E_{po}$ in the cosmic co-moving
  frame of the burst can be written (see, e.g., Pacholczyk, 1970)
  \begin{equation}\label{equation}
  E_{po} = \frac{3he\gamma_{m}^{2} B_{\perp} \Gamma}{4\pi m c 
  }
  \end{equation}
  where $\gamma_{m}$ (which we
  assume is $\gg 1$) 
  is the lower cutoff to the electron power law distribution,
  $N(\gamma)= N_{o}\gamma^{-p}$,
  $\Gamma$ is the bulk Lorentz factor of the emitting source,
   and $B_{\perp}$ is
  the perpendicular component of the magnetic field.  
  To a very rough approximation, we may estimate the radiated energy
  of the GRB as the duration $T$ times the total power emitted per electron,
  integrated over the initial electron energy distribution:
  \begin{equation}
  {\cal{E}}_{\rm rad} \propto T \Gamma^{2}
  N_{o}\int_{\gamma_{m}}^{\infty} \gamma^{-p} d\gamma
  [\frac{4 \sigma_{T} c B^{2} \gamma^{2}}{24 \pi}]
  \propto T N_{o} B^{2}\Gamma^{2}
  \gamma_{m}^{3-p}
 \end{equation}
 where $\sigma_{T}$ is the Thomson
 cross section, and the expression in brackets is the total power radiated by an
 electron of Lorentz factor $\gamma$.  The last proportionality
 comes from performing the integration for
 $p>3$.
 If, for example, in a very simple model
 where $\gamma_{m}$, $T$, and $N_{o}$ were constant from
 burst to burst or independent of $B_{\perp}$ and $\Gamma$,
 then we would find that $E_{po}  \propto {\cal{E}}_{\rm rad}^{1/2} $,
 consistent with our results above. 
However, this ignores many of the subtleties of the GRB
emission.  More detailed models which include information about
the progenitor and environments around the GRB
will predict different
  relationships among the parameters $B$, $T$, 
  $N_{o}$, $\Gamma$, and $\gamma_{min}$
  and - in particular - relate them to the total radiated energy
  of the burst in a more precise way than done above. 
    
It is generally agreed that the GRB radiation is produced by
the action of a blast wave with two basic independent
parameters - the total energy, ${\cal{E}}_{\rm tot}$, and
the bulk Lorentz factor, $\Gamma$.  However, whether
the gamma-ray emission occurs in external or internal shocks
is a matter of dispute.  In the {\em the external blast wave
  model} (see, e.g.,
  Dermer, Chiang, and Boettcher, 1998, and reference cited therein),
  both $\gamma_{m}$ and $B$ will be proportional to the bulk Lorentz
  factor so that $E_{po} \propto \Gamma^{4}$.  The proportionality constant
  depends on details such as the circumburst density, shock compression
  ratio, the fraction of energy that goes into the magnetic field or
  relativistic electrons, but {\em not} on the other basic parameter
  of the blast wave - namely, the total energy ${\cal{E}}_{\rm tot}$.
 On the other hand, the radiated energy ${\cal{E}}_{\rm rad}$ is assumed
 to be proportional to ${\cal{E}}_{\rm tot}$ with the proportionality
 constant fairly independent of $\Gamma$.  Thus, we expect no
  observed correlation between $E_{p}$ and $F_{tot}$, except for the weak
  cosmological contribution.  The strong correlation we see in the data is
  therefore in disagreement with this model.
     
   The above relations are
   more complicated in an {\em internal shock scenario} (Rees \& M\'esz\'aros, 1994,
   Sari \& Piran, 1997),
   because of the presence of additional parameters such as the 
   Lorentz factors of the shocked shells, the shells initial
   separations,  the width of the complex of shells, etc.
   (see Piran, 1999 for a review).   From equations (53), (83), and
   (84) of Piran, one can show that $E_{po} \propto {\cal{E}}_{\rm
   tot}^{1/2}
 \Gamma^{-2}
   \gamma_{int}^{5/2}$, where $\gamma_{int}$ is the Lorentz factor of the 
   internal shock.  If again we assume that $\cal{E_{\rm rad}} \propto
   \cal{E_{\rm tot}}$, with proportionality constant independent of 
   $\gamma_{int}$, $\Gamma$, etc., and that $\cal{E_{\rm rad}}$ and
   $\Gamma$ (or $\gamma_{int}$) are independent parameters of the model,
   then we find $E_{po} \propto {\cal{E}}_{\rm rad}^{1/2}$.  This is consistent
   with the results presented above; thus, we may conclude that the
   analyses presented here favor the internal shock model.

\section{Conclusion}

   From the few bursts with measured redshifts, it is clear that 
 GRBs have a broad luminosity function (Figures 1a and 1b); as a result, direct
 detection of cosmological signatures is difficult, and requires
 a large number of GRBs with known redshifts.  
  Hence, we must continue to rely on statistical studies of GRB properties
  to look for such signatures.  In this paper we analyze the correlation between
  observed burst strengths (peak flux or fluence) and photon spectrum characterized 
  by the peak energy, $E_{p}$, of the $\nu F_{\nu}$ spectrum.  Some or
  all of this correlation could be due to the combined
  effects of spectral redshift and
  source dimming with distance.
  There has been some evidence for such a correlation.
 
 The data we use for these purposes come 
   from the BATSE detector on CGRO.  Our spectral parameters - in particular
   $E_{p}$ - are obtained from
   fits to 16 channel CONT data; the fluences and peak fluxes are taken
   from the 4B catalog (Paciesas et al., 1999).
   In order to get an accurate measure
   of the correlations present in the data, we must be sure to account for
   all types of selection effects present.  In LP99, we discuss the observational
   thresholds on both $E_{p}$ and measures of burst intensity and how
   this might affect correlation studies, and in the Appendix
   of this paper we review the non-parametric techniques that
   we use to account for data
   truncations.
     However, in addition to these truncation
   effects, there are subtle selection effects specific to bursts with
   spectral data.  In particular, the bursts near the instrument's
   threshold generally do not have sufficient data for a
   reliable spectral fit.  Hence,
   the existing  samples are missing many bursts with peak counts $C_{max}$
   near
   the threshold $C_{min}$.  It is difficult to quantify what the effect
   of this will be on our correlation studies.
    To circumvent these additional selection biases, we selected 
   a sub-sample of bright bursts which are fairly
   complete in terms of $C_{max}/C_{min}$.
   This sample has a well defined selection criterion, and hence the
   results we obtain from this sample are more robust and reliable.   
   Applying our methods, we find the following results:
   
   a) We find a strong correlation between peak flux and $E_{p}$ for
   the whole spectral sample, in agreement with earlier results reported
   by Mallozzi et al. (1995, 1998).  However, the sub-sample of bright bursts
   selected in Figure 2a
   shows only a marginal correlation between peak flux and $E_{p}$.
   On the other hand, we find 
   significant correlations between the burst fluence and $E_{p}$
   for both the whole 
   sample as well as the sub-sample shown in Figure 2b.
   The sub-sample, however, shows
   a  much steeper functional dependence 
   %between $E_{p}$ and $F_{tot}$ 
   than
   the whole sample (Figure 3b).  In our subsequent analysis, we
   focus on the total fluence, $F_{tot}$, and $E_{p}$ 
   correlation in the sub-sample, because the truncation effects
   in this sub-sample are better understood, and the investigation
   of the cosmological interpretation
   is simplest in this case. 
   
   b) We quantitatively test to see if our
   $F_{tot}-E_{p}$ correlation can be attributed to cosmological effects alone.
   This, of course, depends on the cosmological model,
   intrinsic distributions of the burst
   parameters, as well as the co-moving rate density of GRBs.  We start
   with the assumption that there are no correlations between intrinsic
   burst parameters, and that cosmological evolution of these parameters is
   minimal.  We present results for an Einstein-de Sitter cosmological
   model, although similar conclusions are reached in other reasonable
   cosmological models we have tested.
   
   c) We find that the observed correlation can be explained by cosmological
   expansion alone if the total radiated energy (in the gamma ray range) is
   constant (standard candle assumption).  This result is independent
   of the GRB rate density, and is fairly insensitive
   to the distribution of the other intrinsic burst parameters.  However,
   as mentioned above, a narrow distribution of radiated energy or luminosity
   is inconsistent with the data from bursts with observed redshifts.
   
   d) For a more plausible, broad luminosity function (e.g., a power law distribution
   $\phi({\cal E}_{\rm rad}) \propto {\cal E}_{\rm rad}^{-\delta}$), 
   we find that neither for a constant co-moving rate density nor for rates
   proportional to the star formation rate, can the correlation be solely
   attributed to cosmological expansion.  The expected correlation is
   essentially ``washed out'' by the broad luminosity function.
   
   e)  This implies either the presence of strong evolutionary effects, or
   an intrinsic correlation between GRB properties.  We find
   that the first possibility is less likely than the second.
   We conclude, then, that the observed correlation between $F_{tot}$
   and $E_{p}$ must be primarily due to an intrinsic correlation between
   the total radiated energy, ${\cal E}_{\rm rad}$, and the rest frame
   peak energy, $E_{po}$.  Using our methods, we find that for broad luminosity
   functions, we can explain the observations if the intrinsic correlation
   obeys a power law, $E_{po} \propto {\cal E}_{\rm rad}^{\eta}$, with
   $0.7 < \eta < 0.4$, depending on the form of the co-moving rate density.
   
   f) This kind of correlation seems to be a natural consequence of 
   optically thin synchrotron
   emission by a power law distribution of electrons with Lorentz
   factors greater than some minimum cutoff, $\gamma_{m}$.  Using the results
   from a detailed modeling of the emission in an external 
   or internal shock model, we find that the internal shock
   model can explain the strong intrinsic correlation more simply.
   
   These results are based on a sub-sample of bursts with a limited range
   of intensity. This range can be expanded and the results can 
   be put on more solid footing by using the complete sample of
   GRBs observed by BATSE. This, however, requires a better
   understanding of the selection effects involved in the entire process
   of detection and spectral fitting.
   These questions will be explored
   in more detail in a future publication.
   
   We would like to thank Rob Preece and Ralph Wijers for useful 
   discussions.  This research was supported by NASA grant NAG5-7874.
    
\appendix{   
\section{Appendix}

\subsection{Estimating the True Correlation of Truncated Data}
We want to examine the correlation of $E_{p}$ 
with some measure of burst intensity - for example, the
total fluence, $F_{tot}$.  However, we
need to account for the fact that $E_{p}$ has
both an upper and lower limit, $E_{p,max}$ and $E_{p,min}$,
 and that $F_{tot}$ has a 
lower limit, $F_{lim}$, to get an accurate estimate of the
true correlation.
We can use the 
$\tau$ statistic to test the degree of 
correlation.

\begin{equation}\label{equation}
\tau = [\frac{\rm positive \  comparisons - negative \  compar
isons}{ \rm 
total \ comparisons}]\times \sigma^{-1}
\end{equation}
   
A positive comparison corresponds to
(a) $E_{p,i} > E_{p,j}$
and $F_{tot,i} > F_{tot,j}$,  {\em or} 
(b) $E_{p,i} < E_{p,j}$
and $F_{tot,i} < F_{tot,j}$.
A negative comparison corresponds to
(a) $E_{p,i} > E_{p,j}$
and $F_{tot,i} < F_{tot,j}$ {\em or} 
(b) $E_{p,i} < E_{p,j}$
and $F_{tot,i} > F_{tot,j}$.
The variable $\sigma$ is the standard deviation of the numerator;
in the usual Kendell's $\tau$ test with no truncation, $\sigma=
[(4N+10)/(9N(N-1))]^{1/2}$, where $N$ is the total number of data points.
Note that when variables suffer from truncation, 
a point can be compared with another only if it
could have been detected in the limiting range of
the other, and vice versa; that is, two points can
be compared when they are in
the limits of one another -
$E_{p,i} \in [E_{p,max,j},E_{p,min,j}]$,
$E_{p,j} \in [E_{p,max,i},E_{p,min,i}]$,
$F_{tot,i} > F_{lim,j}$, {\em and}
$F_{tot,j} > F_{lim,i}$.

Once we get a value for the numerator in equation(A1) above,
which estimates the degree of correlation between
$E_{p}$ and $F_{tot}$, we
must determine the significance of the correlation by finding
the value of $\sigma$.  For the case when variables are truncated,
we cannot simply use the formula for the usual Kendell's $\tau$
test given above.  
To find $\sigma$, we draw N (where N $\sim$ 100 or more)
bootstrap vectors from the observed
distribution of, for example $E_{p}$,
and compute the comparison statistic (numerator
in equation (A1)) for these vectors (Efron and Petrosian, 1999). 
From the distribution of these comparison statistics, we can compute a mean
and standard deviation.
This gives an estimate of the significance
of the correlation (i.e. what we have defined as $\tau$ above).  
For instance, $\tau=0$ implies no correlation while $\tau=1$ implies
a $1\sigma$ correlation.
We can then determine true correlations between parameters
which suffer from truncation.  Once we have an estimate
of the degree of correlation, we can probe the functional
dependence of the correlation.  This is described in more detail below.

\subsection{Recovering the Original Degree of Correlation}
 
 We perform simulations to demonstrate how well this
method can recover the amount of correlation present
in the original (parent) distribution, 
given that we observe a truncated
sample. 
We simulate the spectra of 1000 bursts. 
The parameter $E_{p}$ and the normalization $A$ are
taken from an assumed distribution, while the indices $\alpha$ 
and $\beta$ are held constant.  Note that we are not attempting
to reproduce the actual distributions of the burst parameters here, but
merely want to illustrate how well this method can recover
the parent correlation given some truncation on the data. Of course, the validity of
the method is independent of the choice of burst parameters we use - we present
those simulations in which the truncation has the most dramatic effect.
From knowledge of the spectral parameters, we can calculate the burst
total fluence, $F_{tot}$.  We then can
immediately compute the degree of correlation between $E_{p}$
and  $F_{tot}$, present in the in the {\em original} sample.
We then simulate the limiting fluence $F_{lim}$ of each burst, 
and ask: How many bursts have fluence above their threshold, $F_{lim}$?
 Only a fraction of bursts (in our simulations,
typically half) will make this cut. This is our {\em observed} sample.

We then apply the methods described above to our {\em observed} sample
to determine the degree of correlation between $E_{p}$ and $F_{tot}$, 
given that these variables all suffer from some well determined
truncation.  
 Below, we present the results of two simulations, using the
$\tau$ statistic to measure the degree of correlation.   Note how
well the method recovers the correlation present in the {\em original} sample,
from the {\em observed} distribution and knowledge of the truncation. 
(a) In the first case, 
$E_{p}$ is from a uniform distribution between 100keV
and 2.5MeV, $\alpha=3.0$, $\beta=-5.0$, the
normalization $A$ is from a power law distribution $F(A)\sim A^{0.3}$, and
$F_{lim}=3.0\times 10^{-9}$.
The original,
``observed'', and corrected correlation
results are given
in Table A.1.
In this case, we have {\em no} correlation present
between $E_{p}$ and $F_{tot}$ in the original
sample, but the truncation 
produces an artificial correlation in the observed
sample, as shown in Figure 7a.  Our methods are able to
recover the correlation present in the original sample,
as seen in Table A.1.
\begin{center}
\centerline{\underline{TABLE A.1}}
\centerline{Simulation results for the case when truncation produces a correlation.}
\begin{tabular}{lccccr}  \hline \hline
Variables & Distribution & Number of Bursts & $\tau$ \\ \hline
$F_{tot}$ with $E_{p}$ & Original & 1000    & 0.9$\sigma$ \\
$F_{tot}$ with $E_{p}$ & Observed & 490     & $>5.0\sigma$\\
$F_{tot}$ with $E_{p}$ & Corrected & 490   & 0.6$\sigma$ \\
\hline \hline
\end{tabular}
\end{center}

\begin{figure}[t]
\leavevmode%\centering
\centerline{
\psfig{file=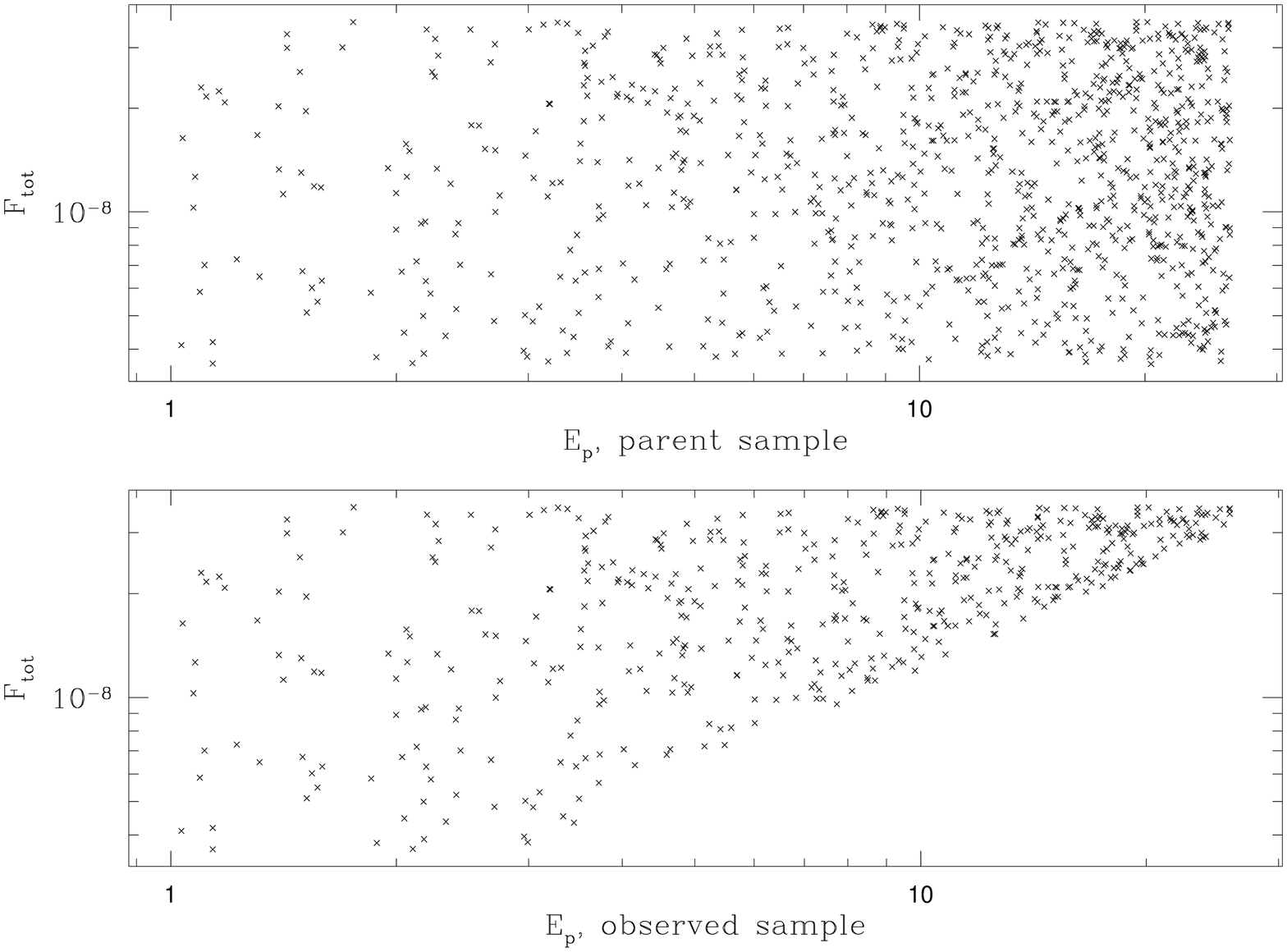,width=0.5\textwidth,height=0.5\textwidth,angle=270}
\psfig{file=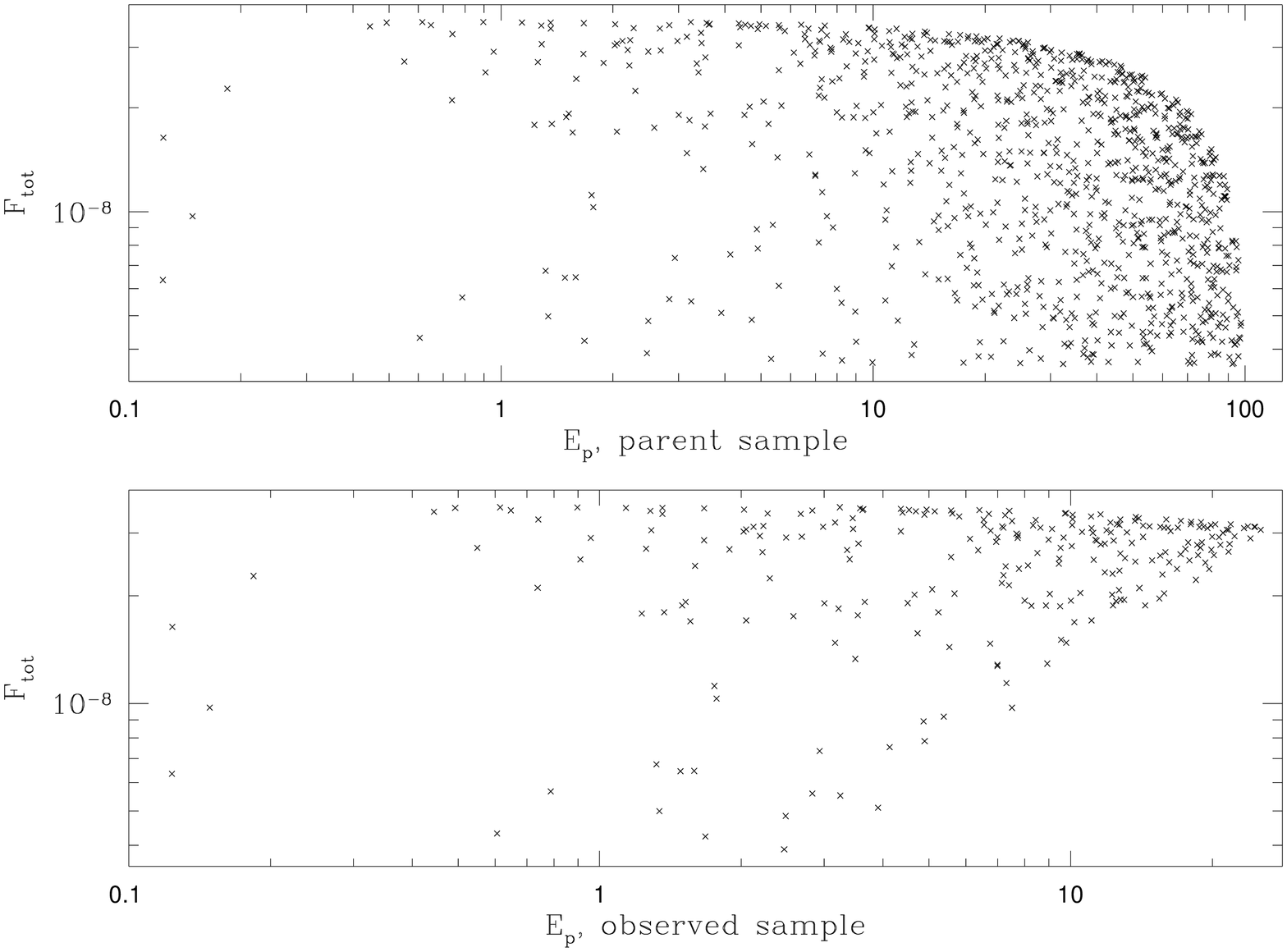,width=0.5\textwidth,height=0.5\textwidth,angle=270}
}
\caption{\, {\bf Left Panels:} (a) Total fluence, $F_{tot}$, vs.
 peak energy, $E_{p}$, for the parent (top panel)
and observed (bottom panel) sample for the first  
simulation in the Appendix.
   Here, the truncation produces a correlation. {\bf Right Panels:}
   (b) 
   Same as Figure 7a, for the second
 simulation given in the Appendix.  Here, the truncation produces
a correlation of the opposite sign than in the original (parent) sample.
}
\end{figure}

 (b) In the second 
case, we
take the normalization from a power law distribution
as in the previous case. The parameter $E_{p} \sim \xi_{i}-(A_{i}/c)^{2}$,
where $\xi$ is some dispersion (a random number between
1 and 100) and $c$ is a constant,
$\alpha$, $\beta$, and $F_{lim}$ have the same values as in
case (a).
Figure 7b shows how the truncation produces a correlation of the opposite
sign from the original correlation.
  Again, the techniques described
above were able to quantify this bias, and
recover the original correlation. The results for this
simulation are given in Table A.2.
\pagebreak
\begin{center}
\centerline{\underline{TABLE A.2}}
\centerline{Simulation results for the case when truncation changes
 the sign of the correlation.}
\begin{tabular}{lccccr}  \hline \hline
Variables & Distribution & Number of Bursts & $\tau$ \\ \hline
$F_{tot}$ with $E_{p}$ & Original & 1000    & (-) $>$5$\sigma$ \\
$F_{tot}$ with $E_{p}$ & Observed & 250     & (+) 2.5$\sigma$\\
$F_{tot}$ with $E_{p}$ & Corrected & 250   &  (-) 4.0$\sigma$ \\
\hline \hline
\end{tabular}
\end{center}

 \subsection{Recovering the Functional Form of the Correlation}

 The method can also be used to recover the {\em functional dependence}
of the correlation present in the original, untruncated sample; this is necessary
in understanding the {\em meaning} of the correlation.
Furthermore, once we know the functional dependence, we can
remove it from our observed sample to obtain an estimate of the frequency
distributions of the relevant variables (see LP99). 
 In this simulation, we begin with a parent sample of 1000
 bursts in which $F_{tot} \propto E_{p}^{1.5}$,
 with some dispersion.
 We truncated the sample in a manner described above, to get
 our observed sample (170 bursts).  We take  our observed sample and find the
correlation between $F_{tot}$ and $E_{p}$, 
accounting for the limits on $E_{p}$ and $F_{tot}$. We then let
\begin{equation}\label{equation}
F_{tot} \longrightarrow F_{tot}' = F_{tot}/q(E_{p}),
\end{equation}
and find the function $q(E_{p})$ such that there is no correlation between
$F_{tot}'$ and $E_{p}$.  Our simulations show that the correlation
is removed when we let $q(E_{p,i}) \propto E_{p}^{\delta}$ where $\delta = 1.51
\pm .1$.
That is, our methods determine the exact functional form of the correlation
present in the parent sample.  Figure 8 shows the degree of correlation
$\tau$ vs. $\delta$ in our simulations.  The correlation is
removed when $\tau = 0$.  The vertical lines mark the $1\sigma$
confidence intervals.  Again, once we have determined the functional form
of the correlation, we can remove the correlation to estimate the true
frequency
distribution of the relevant parameters (LP99).
\begin{figure}[t]
\leavevmode%\centering
\centerline{
\psfig{file=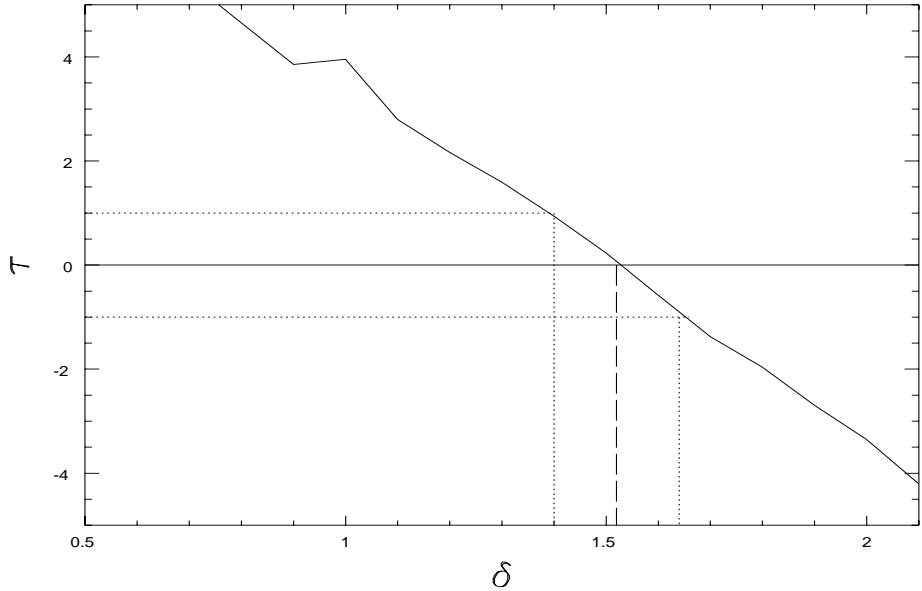,width=0.8\textwidth,height=0.5\textwidth,angle=270}
}
\caption{\,  Significance of the correlation, $\tau$ vs. slope of the power
law, $\delta$, describing the
functional dependence between
$F_{tot}$ and $E_{p}$ for the ``corrected
sample'', as discussed in \S A.3.  We recover from the observed
sample and knowledge of its truncation, the exact functional dependence
present in the parent sample.  
}
\end{figure}

}

\end{document}